\documentstyle[preprint,revtex,eqsecnum]{aps}

\def \eq {\begin{eqnarray}}
\def \eqf {\end{eqnarray}}

\def\pmb#1{\setbox0=\hbox{#1}%
     \copy0\kern-\wd0
     \kern-0.01em\copy0\kern-\wd0
     \kern-0.01em\copy0\kern-\wd0
     \kern0.03em\copy0\kern-\wd0
     \kern0.01em\copy0\kern-\wd0
     \kern-0.04em\raise0.018em\copy0\kern-\wd0
     \kern0.01em\raise0.018em\copy0\kern-\wd0
     \kern0.01em\raise0.018em\copy0\kern-\wd0
     \kern0.01em\raise0.018em\box0}
\def\bbf#1{\mathchoice{\pmb{$\displaystyle #1$}}{\pmb{$\textstyle #1$}}%
     {\pmb{$\scriptstyle #1$}}{\pmb{$\scriptscriptstyle #1$}}}

\begin{document}
\draft
\preprint{LA-UR-92-2771}

\begin{title}

Persistent spin and mass currents and Aharonov-Casher effect\\
\end{title}
\author{A. V. Balatsky}
\begin{instit}
Center for Materials Science and Theoretical Division, T-11,\\
Los Alamos National Laboratory, Los Alamos, NM 87545 and\\
Landau Institute for Theoretical Physics, Moscow, Russia
\end{instit}
\author{B. L. Altshuler}
\begin{instit}
Department of Physics, MIT, Cambridge, MA 02139
\end{instit}

\receipt{\today}
\begin{abstract}
Spin-orbit interaction produces  persistent spin and mass currents in the ring
via the Aharonov-Casher effect. The experiment in $^3He-A_1$ phase, in which
this effect leads to the excitation of mass and spin supercurrent is proposed.

\end{abstract}



The discovery in 1983 of the Berry phase \cite{Berry} brings a new
understanding of different topological effects in quantum mechanics. The
simplest realization of Berry phase is Aharonov-Bohm (AB) effect  of charged
particle in external electromagnetic field. A transparent demonstration of the
AB effect is a persistent current in mesoscopic rings threaded by magnetic
field \cite{persistent}, as well as many other  experiments, which proved the
relevance of
AB effect on transport in multiconnected geometry. More recently it has been
pointed out that there is an analog of AB effect in the presence of the
spin-orbit (SO) interactions, which leads to nontrivial phase shifts and to
topological interference effect of the wave function of a  particle with spin,
which was called Aharonov-Casher (AC) effect \cite{AC}.

There is a major difference between AB and AC effects: the AB effect comes from
the true gauge invariant coupling  $ j_{\mu} A_{\mu}$ between the current  $
j_{\mu}$ and the electromagnetic vector potential $A_{\mu}$ and for that reason
can be observed even in the absence of magnetic field ${\bf B} = curl  {\bf
A}$ in the region
where particles are propagating. In the AC effect, on the other hand, the
phase of the wave function is changed as a result of  SO interaction i.e. as a
coupling of the spin current $ j^{\sigma_i}_{\mu}$ to an effective tensor gauge
potential $ E_{\nu} \epsilon_{\mu \nu i}$ where $ \epsilon_{\mu \nu \lambda}$
is the antisymmetric tensor, and $
{\bf E}$ is the electric field. Therefore nonzero electric field on the way of
the particle is necessary in order to
produce the AC phase shift \cite{Golhaber}.  In the  case of electrons moving
in the atomic electric field SO coupling can be written in a more familiar form
as $\bbf{\sigma} {\bf l}$ with ${\bf l}$ being the orbital momentum of electron
\cite{LL}.

 Inspite this important difference the unifying point of view on both effects
is that they are consequence of the Berry phase acquired by the wave function
of the particle under adiabatic
transport of the particle from some initial state through the set of
intermediate states back to its original configuration.

Here we will implement this point of view and report on the observation that AC
effect can result in   persistent spin and mass currents. The  wave function of
a particle in external magnetic and electric fields and in the presence of SO
interaction will acquire the  spin dependent Berry phase: i) additional phase
of the wave function is given by $ \phi = \Phi_{AB} + \sigma_z \Phi_{AC}$,
where $\Phi_{AB}$ is the AB flux piercing the ring (in case if particles are
charged) and $\Phi_{AC}$ is the AC flux due to AC effect, $\sigma_z = \pm$ is
the spin projection ( we assume spin ${1\over{2}}$ particles). The formerly
doubly degenerate eigenstates and eigenvalues will acquire spin dependent shift
$E_n( \Phi_{AB} + \sigma_z \Phi_{AC}),       \Psi_n(\Phi_{AB} + \sigma_z
\Phi_{AC})$, see also \cite{Entin}; ii) this spin dependent shift will lead to
a persistent spin current:
\eq
j^{\sigma_z}_{\varphi} = -  {c\over{4 \pi R}} Tr {\partial E(\Phi_{AB} +
\sigma_z \Phi_{AC})\over{\partial \Phi_{AC}}} \sigma_z =  -  {c\over{2 \pi R}}
{\partial E\over{\partial \Phi_{AB}}}
\eqf
with $j^{\sigma_z}_{\varphi}$ being the z component of the spin current along
the ring, $\sigma_i$ are the Pauli matricies. In the presence of the net spin
polarization AC effect also leads to mass curent proportional to $n_{\uparrow}
- n_{\downarrow}$.

 Another realization of the Berry phase leading to persistent spin currents was
discussed in \cite{LGB}. Note that the state with persistent spin current does
not violates $P$ (parity) and $T$ (time reversal) invariance, in the contrast
to persistent mass current. Due to this reason persistent spin current can be
excited in the absence of external magnetic flux. Actually for an electron
state  with nonzero angular momentum $\ell$ ( say in an atom) persistent spin
current means nothing but existence of the SO interaction; iii) the AC effect
and persistent spin currents are independent on the charge of the particle and
can be observed for neutral particles , as was done in original observation of
the AC effect for neutrons \cite{Cimmino}; iv)  the  time dependent AC flux
generates effective spin dependent ``electric" field ${\bf E}_{AC}$ via the
Faraday law ${1\over{c}} \partial_t \Phi_{AC} = - \oint {\bf E}_{AC} d{\bf l}$
acting  on neutral as well as on charged spinfull pa!
rticles. This is the local effect,

 independent on the macroscopic phase coherence \cite{Stern}.

The closely related, however different, problem on the effect of random SO
interactions on the electron transport properties in mesoscopic  ring was
considered  by Meir, Gefen and Entin-Wohlman \cite{Entin}. They showed that in
the presence of SO scattering the flux dependence of energy and eigenfunctions
acquires the spin dependent shift $\Phi_{AB} \pm \delta$ where
shift $\delta$ is governed by an averaged over the ring random SO interactions.
Thus the overall effect of SO interaction in this approach is given by the
particualr impurity configuration and independent of external fields. In our
case we will neglect the random SO scattering and consider the AC effect in
external electric field, which can be varied in the direction and magnitude.
This will lead to spin current excitation due to Faraday law for time dependent
AC flux. Interesting realization of this effect will be shown below to take
place in superfluid $^3He$, where external crossed electric and magnetic field
should cause a supercurrent.

Consider first a one dimensional ring of radius $R$. We will describe it using
a tight binding model on a closed chain of $N$ sites separated by a distance $a
= R/N$. An external magnetic field  perpendicular to the plane ${\bf B}||{\bf
e_z}$ (we will use  cylindrical coordinates given by unit vectors ${\bf
e_{\phi}, e_{\rho}, e_z}$) results in the twisted boundary conditions for a
charged particle wave function $\Psi(N) = exp(i 2\pi \Phi_{AB}/\Phi_{0})
\Psi(0)$ with the AB flux $ \Phi_{AB} = B_z \pi R^2$ and $\Phi_{0} = {h
c\over{e}}$. The Hamiltonian of this chain taking into account the SO
interaction and Zeeman splitting $g \mu_B \hat{\bbf{\sigma}}{\bf B}$ ($g$ being
the gyromagnetic ratio) can be written as
\eq
H = - t \sum_{n, \sigma, \sigma'} {\hat \Lambda}(n) |n, \sigma><n+1, \sigma'| +
h.c. +\nonumber\\
 \sum_{n,  \sigma, \sigma'} (\epsilon_n + g\mu_B \hat{\bbf{\sigma}}{\bf B})
|n,\sigma><n,\sigma'|
\eqf
here $n$ labels sites, $\epsilon_n$ denotes the onsite energies, and $ {\hat
\Lambda}_{\sigma \sigma'}(n) = exp( i {\hat A}_{AC}(n, n+1)_{\sigma \sigma'} =
exp(i {g\mu_B \over{\hbar c}} \hat{\bbf{\sigma}} \int_n^{n+1} d{\bf r} \times
{\bf E}({\bf r}))_{\sigma \sigma'}$, where ${\hat A}_{AC}$ is the AC analog of
the vector potential $ \Phi_{AC} = \sum_{n=1}^{n=N-1} A_{AC}(n, n+1)$.

Following the approach of Ref \cite{Entin} we  use the transfer matrix $T_N$
defined as:
\eq
\pmatrix{\psi_{N} \cr
\psi_{N - 1}} = T_N \pmatrix{\psi_1 \cr
\psi_0}
\eqf
where $\psi_n$ is the spinor wave function of the particle in coordinate
representation, and we drop out the spin indexes for simplicity. Using the
structure of the Hamiltonian Eq(2) we can write  $T_N$ as a direct product:
\eq
T_N = \hat{S} \otimes T_N' , T_N' = t^N \pmatrix{ 1  &  0 \cr
		\sum_{i=1}^N (\epsilon_i + g\mu_B \sigma_z)  & 1}
\eqf
\eq
 \hat{S} =  \prod_{i=1}^N \hat{\Lambda}(i)
\eqf
To derive  this equation we had to neglect the noncommutativity of
$[\hat{\Lambda}, g\mu_B B \sigma_z] \sim O(B\cdot E)$, which leads to higher
order in the external fields contributions. As we will see the Berry phase will
be proportional to the circulation of $E_{\rho}$ on the ring; thus taking into
account commutator will lead to higher order terms. Strictly speaking Eqs(4,5)
are valid in the $N \rightarrow \infty$ limit when $\hat{\Lambda}(0) =
\hat{\Lambda}(N)$.

For the slowly varying electric field ${\bf E}$ ($ |\partial_{\phi}{\bf
E}|/|{\bf E}| \  a \ll 1$, where $a$ is a distance between sites) the spin
transfer matrix $\hat{S}$ equals to:
\eq
\hat{S} = \prod_{i=1}^N exp(i {g\mu_B \over{\hbar c}} \hat{\bbf{\sigma}}
\int_i^{i+1} d{\bf r} \times {\bf E}({\bf r}) ) \nonumber \\
\cong exp(i {g\mu_B \over{\hbar c}} \hat{\bbf{\sigma}} \oint d{\bf r} \times
{\bf E}({\bf r}))
\eqf
We neglected the commutator in the exponent in Eq (6) coming from the
Hausdorff's formula $e^A e^B = e^{A + B + 1/2[A,B]}$, since it is proportional
to $\partial_{\phi}{\bf E}$. Only radial component of electric field $E_{\rho}$
contributes to the contour integral in Eq (6). As a result
\eq
{g\mu_B \over{\hbar c}} \oint d{\bf r} \times {\bf E}({\bf r}) =
 - {\bf e_z} 2\pi {\Phi_{AC}\over{\Phi_0}}
\eqf
using Eqs (6,7) the eigenfunctions  equation with twisted boundary conditions
can be written in the compact form
\eq
T_N' \pmatrix{ \psi_1 \cr
		\psi_0} = exp( 2\pi i {\Phi_{AB}\over{\Phi_0}} + 2\pi i
\sigma_z{\Phi_{AC}\over{\Phi_0}})\pmatrix{ \psi_1 \cr
						\psi_0}
\eqf
It is  obvious that in this geometry spin dependent AC flux coming from SO
interactions enters the transfer matrix as a spin dependent  phase.
This equation  allows to find energy eigenvalues and eigenfunctions of the
tight binding Hamiltonian

      Eq (2) if they are known for the bare problem without SO interactions.
Namely,  the energy spectrum and  the wave functions will depend on the
effective flux which is a sum of spin-independent (AB) and spin dependent (AC)
parts: $E_n = E_n(\Phi_{AB} + \sigma_z \Phi_{AC}), \Psi_n =  \Psi_n(\Phi_{AB} +
\sigma_z \Phi_{AC})$, as was mentioned. The energy dependence on the
$\Phi_{AC}$ leads to the persistent spin current $j^{\sigma_z}_{\varphi}$ (see
Eq (1)).  Note again that the spin dependent Berry phase $\Phi_{AC}$  is
nonzero even for neutral  particles. This equivalence of the AC and AB effect
holds only if the spin relaxation is neglected.

It is instructive to estimate the magnitude of the AC flux for realistic
mesoscopic systems. From Eq (7) it follows:
\eq
{\Phi_{AC}\over{\Phi_0}} =  g \mu_B { R E\over{c \hbar}}
\eqf
The relativistic nature of the AC effect is reflected in the ratio of electric
potential on the scale of the size of the ring $eRE$ to the rest energy of the
particle $mc^2$.

If the electric field is time dependent the variation of the  AC flux  results
in the appearance of a spin-dependent driving force due to Faraday law. This
causes the spin current in accord with Ohm's law. Applying Faraday law to spin
$\uparrow \downarrow$ liquids ( assuming $\Phi_{AB} = const$) we find that
particles experience spin dependent force $ - \sigma_z {1\over{c}} \partial_t
\Phi_{AC} = \sigma_z \oint {\bf E}_{AC} d{\bf l}$. Again, in the presence of
net spin polarization this force will cause mass current.  This effect is a
consequence of the Loretz invariance and electrodynamics and is a local
phenomenon, as it has been pointed out previously \cite{Stern}. In order to
excite spin and mass current due to local force  the global phase coherence
along the ring {\it is not required}.

For the ring of the radius $R \sim 10^{-3}$ cm, in the external electric field
$E \sim 10^5$ v/cm Eq(9) gives for particles with gyromagnetic
ratio $g \sim 1$   that ${\Phi_{AC}\over{\Phi_0}} \sim 10^{-3}$  ( see Eq(9)),
that is a tiny effect. On the other hand, in the semiconductors effective
$g$-factor can be two orders of magnitude larger \cite{Stone}.  For these
samples effective flux will be of the order of $10^{-1}\Phi_0$, which makes the
interference effects associated with AC effect in external fields
experimentally observable. Still the experimental observation of the effect in
a real mesoscopic ring does not look as a simple problem. The main difficulties
are due to the screening of the electric field  and the necessity  to work with
considerably strong magnetic fields which prevent  the usage of a SQUID.

Consider  another class of systems with phase coherence  established on the
macroscopic scale -- superfluid $^3He$. In strong enough magnetic field ($B
\sim 10 kG$) superfluid transition is known to  split on two phase transitions
\cite{L}: 1) First atoms with spins along the field are paired -- this is
called $ ^3He-A_1$ phase; 2) After second phase transition atoms of both spin
directions are paired in $ ^3He-A_2$ phase.

 $ ^3He-A_1$ phase is a superfluid with $S_z = +1$ Cooper pairs condensate.
Atoms with spins opposite to the magnetic field are not condensed. In certain
range of temperatures $ \Delta T \sim 6 \cdot 10^{-3} mK/kG \cdot B$ the $
^3He-A_1$ is the only superfluid phase. Thus  the  $ ^3He-A_1$ phase  has
important features: 1) the phase coherence of the of $S_z = +1$ condensate over
the macroscopic distances and 2) total spin polarization of the coherent
subsytem. This  implies that in external electric field condensate will exhibit
AC effect.

We propose the following experiment. Consider $ ^3He-A_1$ phase within a
capacitor with  the plates  lying in the $(xy)$ plane , so that  electric field
has only $z$ component. Let the spins be   polarized  along $y$ axis. In this
case the phase of the condensate wave function acquires the position dependent
shift $\phi(x) = \phi(0) + 2\pi {\Phi_{AC}(x)\over{\Phi_0}}$, with
$\Phi_{AC}(x)$ given by  Eq(7) with integral taken along $x$ axis. This
generates  the condensate flow with  the velocity $v_s = {\hbar\over{2
m_3}}\partial_x \phi(x)$, i.e. supercurrent. For $E \sim 10^5 V/cm$ this
current will be $v_s \simeq 2 \cdot 10^{-7} cm/sec$ ( compare with the critical
velocity in $^3He-A$ $v_c \simeq 0.02 cm/sec$ \cite{L}).  The phase difference
accumulated by the Cooper pair upon transport through the capacitor of length
$L$  will be $\Delta \phi \simeq \ 6 \cdot 10^{-4} L[cm]$.  In order to have a
phase difference  $\Delta \phi \simeq 0.1 \cdot 2\pi$ we need to have a
channel!
 of length $L\sim 1m$. Presently m

ost experiments on $^3He$ are done in containers with characteristic size of
few $cm$.

The supercurrent flow can be detected experimentally since it causes the normal
counterflow due to mass conservation. The latter leads to the heat flow
$\dot{Q}$ opposite to the supercurrent \cite{W}:
\eq
\dot{Q} = (v_s - v_n)(\rho_s/\rho) (TS) (Area)
\eqf
where $v_s$ and $v_n$ are superfluid and normal velocities, $\rho_s$ and $\rho$
stand for superfluid and total densities and $S$ is the $^3He$ entropy per unit
volume. Using Eq (10) we can estimate $\dot{Q}$ at $E \sim 10^5 V/cm$ ( $v_s
\sim 10^{-7} cm/sec$) as $\dot{Q}/area \simeq 10^{-8} erg/sec$.

The ac modification of the same idea looks even more realistically.
We propose the experiment  in the capacitor with  electric field oscillating
with some frequency $\omega$. Such a field will cause oscillating super and
therefore normal currents which apparently can be detected with less
difficulties. The power loss is known \cite{Swift} to be equal to $ P \simeq 8
\pi L \eta v^2_s $, where $L$ is the length of the channel and $\eta  \sim
3\cdot 10^{-2}{g\over{cm sec}}$ is the normal $He^3$ viscosity. The estimation
for the field $E \sim 10^5 V/cm$  gives $P \simeq 10^{-14} erg/sec$. For
example, the frequency dependence of the impedance of the  capacitor can be
measured with the high accuracy. Due to viscous flow of the normal component
the impedance in $^3He-A_1$ will be different from the both normal and any
other superfluid phase : only in $A_1$ phase the AC effect leads to the mass
supercurrent and therefore to the normal component flow.

To conclude , we proposed the new realization of the AC effect, which leads to
persistent spin and mass currents in mesoscopic ring and in the superfluid
$^3He-A_1$ phase. We argue that time dependent electric field in particular
geometry will excite {\it locally} spin current even in the macroscopic samples
via the Faraday law.

We acknowledge useful discussions with A. Leggett, D. Osheroff and G. Swift.
This work was supported by J. R. Oppenheimer fellowship and by Department of
Energy (A.B.), by  Joint Services Electronic Program Contract No DAAL03-89-0001
 (B.A.) and by Advanced Studies Program  of the Center for Materials Science at
LANL.

\end{document}